\documentclass[a4paper,12pt,onecolumn,final,notitlepage,oneside]{article}
\usepackage[cp1251]{inputenc}
\usepackage[T2A]{fontenc}
\usepackage[english]{babel}
\usepackage{indentfirst}
\usepackage{geometry}
\usepackage[dvips]{graphicx}
\usepackage[final]{epsfig}
\usepackage{multicol}
\usepackage{graphicx}
\usepackage{subfigure}
\usepackage{amsmath}
\usepackage{enumerate}
\usepackage{amssymb}
\usepackage{amscd}
\usepackage{hhline}
\usepackage{multirow}
\usepackage{setspace}
\usepackage{makeidx}
\usepackage{textcomp}
\usepackage{amsfonts}
\usepackage{afterpage}
\usepackage{longtable}
\usepackage{cite}
\usepackage{rawfonts}
\usepackage{array}
\usepackage{caption}
 \graphicspath{{pictures/}}

\begin{document}
\title{ The transformation of affine velocity and its application to a rotating disk. }
\author{V. V. Voytik*\thanks {450000, Russia, Ufa, Lenin st., 3, Bashkirian state medical university, department of medical physics with a computer science course, voytik1@yandex.ru}\,\,, N. G. Migranov \footnote{450000, Russia, Ufa, Lenin st., 3, Bashkirian state medical university, department of medical physics with a computer science course, ufangm@yandex.ru}}

\maketitle
\begin{abstract}
 The aim of the article is to find a transformation that links the local affine velocity of a non-rigid body in the laboratory inertial reference frame $ S $ with the centro-affine velocity of motion of this body in the accompanying accelerated frame $ k $. This paper is based on the kinematics of a continuous medium and the generalized Lorentz transformation.
In this paper we show the 3D transformation of velocity  linking the reference system $ S $ and the reference system $ k $, which moves without rotation. 
Wherein the motion of various points of the rigid system $ k $ is inhomogeneous. Using these formulas, we obtain the desired direct and inverse transformation of the local affine velocity. Important special cases of this transformation are considered. They are the motion of particles in a uniform force field and the precession of Thomas.
As an example of using the transformation of affine velocity in $ S $, accelerated rotation of the disk was considered and the local angular velocity and the magnitude of the deformation of its points were calculated. Wherein, the calculated stretching coefficient is consistent with the known one, and the formula found for the angular velocity is more general than the earlier result obtained for uniform rotation of the disk.
\end{abstract}



\begin{flushleft}
{\bf{ Introduction }}
\end{flushleft}


It is well known that the existence of the limiting velocity of propagation of interactions $ c $ does not allow the existence of ideally rigid bodies. Indeed, if one part of the body is set in motion, then the part that is located at a distance of $ l $ from it will start moving not earlier than in time $ l / c $. All this time the body will be deformed \cite{42}.

It would seem that this conclusion can be circumvented if all parts of the body are equipped with programmable engines. The engine should turn on at the right time and act on its body part with some force, so as not to change its initial proper position. However, even in this case, for a variable angular velocity of the starting point of the body, tangential distances (tangential to a circle around the angular velocity vector) are not preserved \cite[paragraph 84]{43}. Thus, in the general case in relativistic physics (as opposed to classical) global (including the entire space) rigid non-inertial reference systems do not exist.

If the proper dimensions of the frame of reference are small enough, then such a frame of reference is called the local system \cite {40}. Note that the preservation of the size of the local body depends on the frame of reference that the observer chooses. If an arbitrary local reference system $ a $ moves in such a way that it retains its dimensions relative to the laboratory system, then it is a non-rigid system. This was shown, for example, in \cite {46} and \cite {45}. These articles have been reviewed, respectively, the rectilinear motion of the local frame of reference $ a $ and the rotation of $ a $ along a circle with angular acceleration.

One can neglect the proper tangential deformation of the local rigid system $ r $. However, with a non-inertial movement of $ r $, its dimensions in the laboratory frame reference $ S $ change, it looks non-rigid \cite {42}. This circumstance is easy to show. Considering the accelerated rod it is obvious that the higher its speed, the more susceptible it is to the reduction of Lorentz. Consequently, the rod moves as a non-rigid body \cite {47}. Similarly, it affects the stiffness property and the rotation of the reference system $ r $. Indeed, let the trajectory of a point of a rotating coordinate system relative to the accompanying inertial reference system be a circle. With respect to another inertial frame reference, this trajectory is not closed. Let us ask ourselves what this trajectory will be if in the laboratory frame of reference $ S $ to exclude the systematic drift of the leading center of the trajectory? This figure due to the Lorentz contraction must be an ellipse, in which, as is known, the distance from the point of the ellipse to its center depends on the specific direction. Therefore, it is impossible to speak of an exceptionally rigid rotation with respect to the $ S $ of the reference system $ r $ even if its beginning moves uniformly (not to mention the non-uniform case). The same conclusion is true for, for example, the Thomas precession. The Thomas precession here is understood to mean the rotation of the coordinate axes of the accelerated reference frame with a known frequency (the value of which is debatable) in the laboratory inertial reference frame \cite [Appendix D] {16}. If we consider that the frame reference may not necessarily be rigid, then the conclusion about the complexity of the movement of the points of its coordinate system is obvious.

For this reason, in the relativistic theory one cannot do without considering non-rigid motion both with respect to the accompanying local rigid frame $ r $ and with respect to $ S $. Therefore, it is necessary to clarify the important connection between the centro-affine \cite {48} non-rigid motion inside $ r $ and the same motion with respect to $ S $. Recall that an affine transformation is a mapping of space into itself, in which parallel lines transform into parallel lines, intersecting - into intersecting ones. The centroaffine transform preserves the  coordinate origin.

From a mathematical point of view, local centroaffine motion reduces to a system of homogeneous differential equations of the first order (see below the formula \eqref {2.25}). In these equations, the rate of change of the unknown radius vector is proportional with the variable tensor coefficient to the desired vector itself. In the well-known manuals, the solution of such a system is given only for a constant coefficient \cite {49}. In the general case, for a homogeneous equation, a closed method of solution does not exist. For this reason, the well-known radius vector in $ r $ as an explicit function of time is an exception. As a rule, only a tensor coefficient is known at a given time.
It is proposed to call it affine velocity.

Affine velocities in reference systems $ r $ and $ S $ are connected by some local linear transformation. It is required to calculate it. For definiteness, as the system $ r $, we choose a non-rotating reference frame $ k $, which currently accompanies and coincides with $ r $. In this case, it is obvious that the matrix of the desired transformation depends only on the velocity and acceleration $ k $.

If there is no motion in the $ k $ system, then the affine velocity of the rod that belongs to $ k $ is zero and this rod is rigid. However, in $ S $ its affine speed is nonzero. She was found in \cite {16}. The general case of non-rigid motion in the $ k $ system is considered in this paper. 

Non-rigid movement is considered classical continuum mechanics. Therefore, the article is based on two theories: both the special theory of relativity (STR) and the classical     
kinematics of continuum (CKC) \cite {60}. Wherein, the CKC competence concerns some statements in the accompanying frame of reference and similar statements in the laboratory frame reference. At the same time, STR uses conclusions concerning the transition between the laboratory frame reference  and the accompanying frame of reference. Therefore, in this case, these theories do not contradict each other.

The aim of the article is to study and apply the transformation of an arbitrary three-dimensional affine velocity tensor. We believe that solving this problem will allow us to correctly understand and calculate the rotational speed and deformation of an arbitrary local non-rigid body.
 
Let us briefly discuss the structure of the article.
In the next section, we review some preliminary information about research methods that are important for further discussion. In the section on the kinematics of a continuous medium, the tensor of the affine velocity of a general form is considered and the physical meaning of its components is clarified. In the paragraph about the mathematical basis of special relativity, the generalized Lorentz transformation into a non-inertial reference frame $ s $ is given, which is oriented in a special way and its characteristics are calculated. After that, a section of the results is presented.
First, a three-dimensional velocity transformation is found that connects the systems $ S $ and $ s $, $ S $ and $ k $. Then we find the velocity parameter $\mathbf {v}(t)$, which describes the motion of the system $ s $ and $ k $ and enters the generalized Lorentz transformation as a function of the proper  coordinate. After this, we assume that the strain rate in the accompanying non-rotating reference frame $ k $ is small. This circumstance will allow decomposing the velocity function in $ S $ in the first order in its proper coordinate. Further, in Section 3 we give the transformation of the affine velocity matrix from the frame  $ k $ to $ S $. Similarly, in paragraph 4, the inverse affine velocity transformation is written.
At the end of the results section, we consider the application of the formulas obtained to rigid accelerated rotation of the body in a laboratory frame reference.

The material that is described in this paper is available to undergraduate students.

\subsubsection*{ Research methods.}

\subsubsection*{1. The method, that uses Euler’s point of view on continuum kinematics.} 

In continuum mechanics, there are two basic equivalent formulations.

One of these methods uses the Lagrange point of view on the movement of a continuous medium. According to this point of view, each point of the medium is assigned its individual coordinates \cite{60}. These coordinates do not change during the movement time. This means that all points of the medium are at rest relative to the moving accompanying coordinate system. However, the coordinate system itself is curvilinear and deformed. The basic quantity from the Lagrange point of view is the metric tensor of the accompanying coordinate system as a function of time.

This article is based on another formulation that uses Euler’s point of view. According to Euler, a rigid coordinate system is used as the reference system of the observer. In this case, the observer is interested in what happens at a given point in space. The main value from the point of view of Euler is the Cartesian coordinate of the medium point as a function of time. Take as a frame reference $ k $ a rectangular coordinate system that accompanies one of the medium points. Let we know the velocities
medium points $u^{\alpha}$ as a function of time $ t $ and coordinates $r^{\alpha} $. 
\begin{equation} \label{25} 
\frac{dr^{\alpha}}{dt} =u^{\alpha}(r^{\beta},t)\,,                                                          
\end{equation} 
The velocity of the point $r^{\alpha} = 0 $ is zero: $u ^{\alpha} = 0$. Therefore, the function $u^{\alpha} (r^{\beta}, t) $ can be expanded in a series in powers of $r^{\beta}$. By limiting ourselves to the first degree, we obtain a system of homogeneous linear differential equations of the first order
\begin{equation} \label{2.25} 
\frac{dr^{\alpha }}{dt} =\omega ^{\alpha\beta } r^{\beta }\,.                                                         
\end{equation} 
Solving the \eqref {2.25} system, we can find the dependence of the coordinates $r^{\alpha} $ on time and on three arbitrary constants. These constants are the individual coordinates of points according to the Lagrange formulation.
  
Similar movement of points occurs in the laboratory frame of reference. The affine motion of a non-rigid rod in the laboratory reference frame obeys the equation
\begin{equation} \label{255} 
U^{\alpha} =V^{\alpha }+\Omega^{\alpha \beta } L^{\beta}\,,
 \end{equation} 
where $\mathbf{U} $ is the velocity of the rod end, $\mathbf{V} $ is the velocity of the rod origin, $\mathbf{L} $ is its length in the laboratory frame of reference.
Considering that
\begin{equation} \label{255.2} 
U^{\alpha } -V^{\alpha }= \frac{dL^{\alpha}}{dT}\,,
  \end{equation}	
	the equality \eqref {255} can be written as 
\begin{equation} \label{255.1} 
\frac{dL^{\alpha}}{dT}=\Omega^{\alpha \beta } L^{\beta}\,.
  \end{equation}		
 
The value of $\omega^{\alpha \beta}$ is sometimes called the generalized angular velocity of a rod. However, this name is not correct. For example, for affine stretching $\omega^{\alpha \beta} $ is nonzero, but this transformation is not a rotation, so that the name “angular velocity” is not applicable. It would be more correct to call the tensor $\omega^{\alpha\beta} $ its proper affine or affinity velocity.
The question arises about the meaning of the components of this tensor. It is convenient to decompose the tensor of affine velocity into the symmetric part of $s^{\alpha\beta}$ and antisymmetric, where the antisymmetric part of the tensor $\omega^{\alpha \beta}$ is dual to some vector $\omega^{\gamma}$
\begin{equation} \label{2.39} 
\omega ^{\alpha \beta } =s^{\alpha \beta } +e^{\alpha \gamma \beta } \omega ^{\gamma} \,.                                           
\end{equation} 
The values of $s^{\alpha\beta}$, $\omega^{\gamma}$ are
\begin{equation} \label{2.40} 
\omega ^{\alpha } =\frac{1}{4}\, e^{\alpha \mu \nu } \left(\omega ^{\nu \mu } -\omega ^{\mu \nu } \right)\,,                                            
\end{equation} 
\begin{equation} \label{2.41} 
s^{\alpha \beta } =\frac{1}{2} \left(\omega ^{\alpha \beta } +\omega ^{\beta \alpha } \right)\,.                                                
\end{equation} 
The tensor $s^{\alpha\beta}$ is called the strain rate tensor. The physical meaning of these quantities is determined by substituting \eqref {2.39} into \eqref {2.25}. Let's go to the reference frame, which rotating with an angular velocity $\omega^{\gamma}$. In such a frame of reference, the equation \eqref {2.25} takes the form 
\begin{equation} \label{25.1} 
u^{\alpha}=s^{\alpha \beta}r^{\beta}\,.                                                     
\end{equation}
We now choose a coordinate system in which the tensor $s^{\alpha \beta}$ takes a diagonal form. In such a coordinate system
\begin{equation} \label{2.28} 
s^{\alpha \beta}=s^{(\alpha)}\delta^{\alpha\beta}\,.
\end{equation}
The principal values $s^{(\alpha)}$ of this tensor along the $\alpha $ axis are determined from the equation
\begin{equation} \label{27} 
\left|\,s^{\alpha \beta}-s^{(\alpha)}\delta^{\alpha\beta}\right|=0\,,
\end{equation}
where by index, which standing in the bracket,  summation  is not.
The unit vectors $n^\alpha$ of the new coordinate system are determined from the equation
\begin{equation} \label{28} 
s^{\alpha \beta}n^{\beta}=s^{(\alpha)}n^{\alpha}\,.
\end{equation}
The equation \eqref {25.1} will have the following form
\begin{equation} \label{25.2} 
u^{\alpha}=\frac{\partial\, r^{\alpha}}{\partial t}=s^{(\,\alpha) }r^{\alpha}\,.                                                     
\end{equation}
The solution of a differential equation \eqref {25.2} by the method of separation of variables leads to the expression
\begin{equation} \label{25.3} 
\frac{r^{\alpha}}{r^{\alpha}_0 }=\exp{\,\,\int\limits^{t}_0 s^{(\,\alpha)}dt}\,.                                                 \end{equation}
On the other hand, the ratio $r^{\alpha}/r^{\alpha}_0$ is the coefficient of relative elongation of the $ \alpha $ axis. Therefore, if the affine velocity tensor has a general view, it is wrong to speak only about rotation. Knowing the affine velocity tensor $\omega^{\alpha\beta}$, one can find both the principal vectors and the principal values of the strain tensor along them, and the angular velocity of rotation of the main axes.
In this case, $ \boldsymbol {\omega} $ is the angular velocity of rotation of the principal axes of the tensor $s^{\alpha\beta}$, and the principal value of this tensor is the velocity of relative elongation of the corresponding principal axis.

\subsubsection*{2. The method of generalized Lorentz transformation.} 

When solving problems in the SR, two methods for the consideration of noninertial reference frames are also possible.

One of them (historically the first) completely ignores non-inertial reference frames. To simulate a noninertial reference frame, the method of inertial comoving reference frames (ICRF method) is used, which goes back to A. Einstein.
 The essence of the method is that the non-inertial reference frame $r$ at two successive moment of time is replaced by two inertial reference frames $i$ and $i'$, which at these times instantly coincide and accompany the non-inertial frame $r$. 
Such modeling is used to applied Lorentz transformations between the laboratory frame and the $i$ and $i'$ frames. In the classic textbook \cite{36}, using this method, two frames $ i $ and $i'$ were actually considered, which are equally oriented with the $S$ laboratory frame and move with the velocity $\mathbf {v} $ and $\mathbf{v}+\Delta\mathbf{v}$ respectively. 
The author \cite{36} was interested in the calculation of the characteristics of the reference frame, which preserves its orientation in the laboratory frame. It turned out that for close times $ t $ and $t+\Delta t$ the frame $i'$ is not only moving at a speed of $\delta \mathbf{v}$ relative to $i$, but is also rotated by the angle $\delta \boldsymbol{\varphi}$ relative to $i$.
Boost $i'$ is equals \cite [p. 551] {36}

\begin{equation}\label{4.5} 
	\delta \mathbf{v}=\gamma ^2 \Delta\mathbf{v}_{\|}+\gamma \Delta\mathbf{v}_{\bot}\,,
\end{equation}
where $\Delta\mathbf{v}_{\|}$ and $\Delta\mathbf{v}_{\bot}$ is the component of $\delta\mathbf{v}$, which is parallel and perpendicular to the velocity  $\mathbf{v}$ respectively. The formula \eqref{4.5} can be rewritten as
\begin{equation}\label{a1.9} 
	\delta\mathbf{v}=\frac{\Delta \mathbf{v}}{\sqrt{1-v^2}}+\frac{1-\sqrt{1-v^2}}{v^2(1-v^2)}(\mathbf{v}\Delta \mathbf{v})\mathbf{v}
\end{equation}
This acceleration transformation was obtained in \cite {53}.

Angle of rotation is
\begin{equation}\label{a1.10} 
	\delta \boldsymbol{\varphi}=\frac{1-\sqrt{1-v^{2} } }{v^{2} \sqrt{1-v^{2} } }\,\, \mathbf{v}\times \Delta \mathbf{v}\,.
\end{equation}
This rotation by the angle $\delta\boldsymbol{\varphi}$ in the time $\Delta t$ is obtained by successively performing two boosts \cite{36}. This rotation is called the Wigner rotation.
 
In addition to this method, there is another standard method for studying non-inertial reference frames - the method of the generalized Lorentz transformation (or the Lorentz – M{\o}ller – Nelson (LMN) transformation). The generalized Lorentz transformation connects the laboratory inertial reference frame $ S $ and the non-inertial rigid frame $ s $, whose axes move without rotation \footnote{Let us understand the concept of the same orientation of the axes of the laboratory system and the moving reference system $ s $. This question is important because in a number of papers the orientation of the moving system is compared with the orientation of $ S $. In the event that the velocity $ s $ is directed along one of the axes $ S $, then in any sense the axes of the system $ s $ coincide with the axes $ S $ and there are no problems. But, if the direction of the velocity is any other, then the rectangular coordinate axes of the reference system $ s $ are generally not rectangular from the point of view of the observer in the $ S $ system (M{\o}ller). Therefore, the words "without rotation" \ should be understood conventionally, in the sense that the vector spaces of the reference systems $ s $ and $ S $ connected by the generalized Lorentz transformation \eqref {1.1}, \eqref {1.2} are considered to have the same "orientation". The actual orientation of the orts of the systems $ S $ and $ s $ is different and depends on the speed $ s $. Another question is closely connected with this question: can the angle of rotation of the coordinate axes of the moving reference frame $ s $ be attributed to the axes of the laboratory system? We believe that such an operation is incorrect. Therefore, the calculation of the frequency of the Thomas precession in this way is wrong.} relative to $S$. It was opened by the joint efforts of a group of great researchers and took a long time. Therefore, the history of the issue is extremely rich and far from being exhausted by the links given below. The pioneer of this transformation for the case when the frame $s$ moves inertially became in 1899, as is known, Hendrik Lorenz. In 1943, the Danish physicist Christian M{\o}ller succeeded in writing a transformation into a noninertial reference frame $s$ \cite[formulas (64), (66)]{10} that straightforwardly moving in an arbitrary manner along the $X$ axis. In the final form for a rigid reference frame $s$ moving arbitrarily without proper rotation, he formulated it in 1952 in the monograph \cite[sections 4.14, 8.15 in the Russian edition]{8} (see also \cite{3}, \cite [paragraph 2]{41}, \cite{44}, \cite [paragraph 2]{50}, \cite[paragraph 2.2]{51}).

We follow the article \cite{3}, in which the generalized Lorentz transformation is given in three-dimensional form. In the laboratory inertial reference frame, the square of the interval $ds$ is equal (hereafter $(c = 1)$)
\begin{equation}\label{1.5} 
	ds^2=dT^2-d\mathbf {R}^2\,.
\end{equation}
The generalized Lorentz transformation from the laboratory inertial reference frame $S$: $(T,\mathbf{R})$ to the reference frame $s$: $ (t,\mathbf{r})$ is a transformation
\begin{equation} \label{1.1} 
T=\frac{\mathbf{vr}}{\sqrt{1-v^{2} } } +\int \limits_{0}^{t}\frac{dt}{\sqrt{1-v^{2} } }   
\end{equation} 
\begin{equation} \label{1.2} 
\mathbf{R}=\mathbf{r}+\frac{1-\sqrt{1-v^{2} } }{v^{2} \sqrt{1-v^{2} } } \mathbf{(vr)v}+\int\limits _{0}^{t}\frac{\mathbf{v}dt}{\sqrt{1-v^{2} } }.                                
\end{equation} 

By differentiating these equalities, we can obtain the relations \eqref{2.5}, \eqref{2.6} (see Appendix 1). After substituting them into \eqref{1.5}, it is easy to show that the mathematical form $ds^2$ is converted to
\begin{equation}\label{1.6} 
	ds^2=[(1+\mathbf {Wr})^2-(\mathbf {\Omega \times r})^2]dt^2-2(\mathbf {\Omega \times r})d\mathbf {r}dt-d\mathbf {r}^2
\end{equation}
 where                 
\begin{equation} \label{3} 
\mathbf{W}=\frac{\dot{\mathbf{v}}}{\sqrt{1-v^{2} } } +\frac{1-\sqrt{1-v^{2} } }{v^{2} (1-v^{2} )} \mathbf{(\dot{v}v)v} 
\end{equation} 
and
\begin{equation} \label{4} 
\mathbf{\Omega} =\mathbf{\Omega}_{W} =\frac{1-\sqrt{1-v^{2} } }{v^{2} \sqrt{1-v^{2} } }\,\, \mathbf{v\times \dot{v}}, \,\,\,\,\,      \mathbf{\dot{v}}=\frac{d\mathbf{v}}{dt}\,.  
\end{equation} 
Let us write down separately $\boldsymbol{\Omega}_ {W}$ expressed in terms related only to the laboratory reference frame
\begin{equation}\label{1.15} 
	 \boldsymbol{\Omega}_{W}=\frac{1-\sqrt{1-v^{2} } }{v^{2} (1-v^{2} ) }\,\, \mathbf{v}\times  \frac{d\mathbf{v}}{dT}=\frac{\gamma ^3}{\gamma +1}\,\,\mathbf{v}\times  \frac{d\mathbf{v}}{dT}\,.
\end{equation}

The square of the interval $ds$ \eqref{1.6} is the interval of the accelerated and rotating reference frame \cite [section 13.6, p. 404]{11}, \cite{41}, \cite{50}. The small difference in the proper metric \cite{11} and \cite{41}, \cite{50} is due to the fact that \cite{11} shows the metric with an accuracy of the first degree in its proper coordinate. Consequently, the reference frame $s$ has  proper acceleration \eqref{3} and  proper angular velocity \eqref{4}.

We now divide \eqref{a1.9} and \eqref{a1.10} by $\Delta t$. Then it is easy to obtain, respectively, the formulas \eqref{3} and \eqref{4}. Thus, the method of comoving inertial reference frames and the LMN transformation method are physically equivalent, since they lead to the same results.

 The disadvantage of ICRF method is that the only moving noninertial reference frame $s$ is already assumed to be a lots of  inertial reference frames. Obviously, however, such a replacement requires additional justification, if only because of the presence of inertial forces in $s$ and their absence in $i$, $i'$. In addition, such reference frames, we must consider at least two. Considering also the laboratory frame there will be three of them already. This greatly complicates the consideration of the physical problem.

In our opinion, it is methodically more correct to use the generalized Lorentz transformation. Indeed, the same characteristics of the  frame $s$ can be obtained using one generalized Lorentz transformation and thereby passing into just one noninertial reference frame. Such a conclusion significantly reduces computation, saves time, is more obvious and physically easier. Therefore, this article is based on the LMN transformation method.

\subsubsection*{Results}
\label{sec}

\subsubsection*{1. The velocity transformation from a lab system to a concomitant nonrotating frame and vice versa.}

Consider the motion of a rod relative to the accelerated reference frame $ k $, which accompanies $s$ and does not have its proper rotation. In this case, the $k$ axes at the moment coincide with the $s$ axes. It turns out that in this case the corresponding formulas are slightly simpler.
 The speed $\mathbf{U}$ of the end of a rod in the laboratory frame $S$ is related to the velocity $\mathbf{u}$ in the frame $k$ by equations
 	\begin{equation} \label{a2.23} 
\mathbf{U}=\frac{(1+\mathbf{Wr)v}+\sqrt{1-v^{2} }\,\,\mathbf{u} +\frac{1-\sqrt{1-v^{2} } }{v^{2} } (\mathbf{vu} )\; \mathbf{v}}{1+\mathbf{Wr+vu} }  
\end{equation} 
\begin{equation} \label{a2.24} 
\mathbf{u} =\frac{(1+\mathbf{Wr})\; \left[\sqrt{1-v^{2} } \,\,\mathbf{U-v}+\frac{1-\sqrt{1-v^{2} } }{v^{2} } \mathbf{(vU)v}\right]}{1-\mathbf{vU}}  
\end{equation} 
The calculation is given in Appendix 1.

Note that if we substitute into \eqref{a2.23} $\mathbf{u} = 0$, we get $\mathbf{U = v}$.

\subsubsection*{2. Parameter $\mathbf{v}$ as a function of its proper coordinate.}

Let us show that the parameter $\mathbf{v}$ is a function of the velocity  $\mathbf{V}$ origin  frame  $k$, its laboratory acceleration $\mathbf{A}$ and the coordinates $\mathbf{r}$.

It is easy to see that the LMN transformation parameter — the function $\mathbf{v}$ — is not dependent on the time $T$ of the laboratory frame, but on the time $t$ of its proper frame. This circumstance is a characteristic property of the generalized Lorentz transformation. Instead of time $t$ we introduce a new variable according to the condition
\begin{equation}\label{a2.25} 
	\int\limits _{0}^{t}\frac{\mathbf{v}dt}{\sqrt{1-v^{2} } }=\theta
\end{equation}
Knowing the function $\mathbf{v}(t)$, we also know the function $\mathbf{v}(\theta)$. According to the equation \eqref{1.1}
\begin{equation}\label{a2.26} 
\theta=T-\Delta T\,,
\end{equation}
where
\begin{equation}\label{a2.70} 
\Delta T=\frac{\mathbf{vr}}{\sqrt{1-v^2}}\,.
\end{equation}
We decompose $\mathbf{v}(\theta)$ in powers of $\Delta T$ according to \eqref{a2.26}, \eqref {a2.70}. If confining the first degree we get
\begin{equation} \label{a2.15} 
\mathbf{v=V}-\frac{\mathbf{(Vr)A}}{\sqrt{1-V^2}}\,, .                   
\end{equation}
Another representation of the formula \eqref{a2.15} through the proper acceleration of $\mathbf {W}$ is
\begin{equation} \label{a2.20} 
\mathbf{v=V}-\sqrt{1-V^{2} }\mathbf{(Vr)W}+\frac{\sqrt{1-V^{2} } (1-\sqrt{1-V^{2} } )}{V^{2} }\mathbf{(Vr)(VW)V} \,.
\end{equation}

\subsubsection*{3. The transformation of laboratory affine velocity into the comoving  nonrotating reference frame.}

So, let the end of the rod in the laboratory reference frame  move according to \eqref{255}. The movement of this point relative to the reference frame $ k $ is interesting. Given the equalities \eqref {2.25} and \eqref {a2.24}, it can be found that
\[\omega ^{\alpha \beta } =\frac{\Omega ^{\alpha \beta } }{\sqrt{1-V^{2} } } +\frac{1-\sqrt{1-V^{2} } }{V^{2} (1-V^{2} )}\, \Omega ^{\gamma \beta } V^{\alpha } V^{\gamma } -\frac{(1-\sqrt{1-V^{2} } )^{2} }{V^{4} (1-V^{2} )} (\Omega ^{\gamma \mu } V^{\gamma } V^{\mu } )V^{\alpha } V^{\beta } -\] 
\begin{equation} \label{52} 
-\frac{1-\sqrt{1-V^{2} } }{V^{2} \sqrt{1-V^{2} } }\, \Omega ^{\alpha \gamma } V^{\gamma } V^{\beta } +\frac{1-\sqrt{1-V^{2} } }{V^{2} \sqrt{1-V^{2} } ^{\,3} }\, (\mathbf{AV)}\, V^{\alpha } V^{\beta } +\frac{V^{\beta } A^{\alpha } }{1-V^{2} }  \,.
\end{equation} 
The symmetric and antisymmetric parts of the tensor $ \omega ^ {\alpha\beta} $ have the form
\[s^{\alpha \beta } =\frac{S^{\alpha \beta } }{\sqrt{1-V^{2} } } +\frac{(1-\sqrt{1-V^{2} } )^{2} }{2V^{2} (1-V^{2} )}\, V^{\gamma } (S^{\alpha \gamma } V^{\beta } +S^{\beta \gamma } V^{\alpha } )-\frac{(1-\sqrt{1-V^{2} } )^{2} }{V^{4} (1-V^{2} )} \,(S^{\gamma \mu } V^{\gamma } V^{\mu } )V^{\alpha } V^{\beta } +\] 
\begin{equation} \label{56} 
+\frac{(A^{\alpha } -e^{\alpha \lambda \gamma } \Omega ^{\lambda } V^{\gamma } )V^{\beta } +V^{\alpha } (A^{\beta } -e^{\beta \lambda \gamma } \Omega ^{\lambda } V^{\gamma } )}{2(1-V^{2} )} +\frac{1-\sqrt{1-V^{2} } }{V^{2} \sqrt{1-V^{2} } ^{\,\,3} } \,(\mathbf{VA})V^{\alpha } V^{\beta }\,,  
\end{equation} 
\begin{equation} \label{55} 
\omega ^{\alpha } =\frac{2-V^2 }{2(1-V^{2}) }\,\Omega ^{\alpha } -\frac{(1-\sqrt{1-V^{2} } )^{2} }{2V^2 (1-V^2)}\,\,  (\mathbf{\Omega V}) V^{\alpha} +\frac{e^{\alpha \mu \nu } V^{\gamma } V^{\nu } S^{\gamma \mu } }{2(1-V^{2} )}+\frac{e^{\alpha \mu \nu } V^{\mu } A^{\nu } }{2(1-V^{2} )}  \,.
\end{equation} 
Here $V^{\alpha}$ is the velocity of the frame origin $k$, $ \Omega ^ {\alpha \beta} $ is some matrix of affine velocity in the laboratory frame, $ \mathbf {\Omega} $ and $ S ^ {\alpha \beta} $ - is the angular velocity  of the principal axes and the strain rate tensor in the laboratory frame, respectively, $ \omega ^ {\alpha \beta} $ - some matrix of affine velocity in a nonrotating frame $k$, $ \boldsymbol {\omega} $ and $ s ^ {\alpha \beta} $ are respectively the angular velocity of the principal axes and the strain rate tensor in the frame $k$.
These calculations are performed in Appendix 3.

We now consider the important special case of the \ eqref {52} transformation.
Let the local reference frame move in such a way that the distance vector between any two points measured in $ S $ was constant during the movement. This requirement means that the affine velocity for this motion is zero ($\Omega^{\alpha \beta} = 0$). However, in the comoving frame, the affine velocity will be different from zero.
If $ \Omega ^ {\alpha \beta} = 0 $, then from the equality \eqref {52} it follows that
\begin{equation} \label{52.1} 
\omega ^{\alpha \beta } =\frac{1-\sqrt{1-V^{2} } }{V^{2} \sqrt{1-V^{2} } ^{\,3} } (\mathbf{VA)}\, V^{\alpha } V^{\beta } +\frac{V^{\beta } A^{\alpha } }{1-V^{2} } \,.
\end{equation} 
and
\begin{equation} \label{55.1} 
\boldsymbol{\omega} =\frac{\mathbf{V\times A} }{2(1-V^{2} )}  \,,
\end{equation} 
\begin{equation} \label{56.1} 
s^{\alpha \beta }=\frac{V^{\beta } A^{\alpha }+V^{\alpha } A^{\beta }}{2(1-V^{2} )} +\frac{1-\sqrt{1-V^{2} } }{V^{2} \sqrt{1-V^{2} } ^{\,\,3} } \,(\mathbf{VA})V^{\alpha } V^{\beta }\,.  
\end{equation} 
Let's find the principal axes and principal values  of $s^{\alpha \beta}$. Choose the original coordinate system so that axis 1 passes in the direction of the velocity vector $ \mathbf {V} $. The axis 2 is in the plane in which the vectors $ \mathbf {A} $ and $ \mathbf {V} $ lie. Let the angle between these vectors be $ \alpha $ (Figure 1).
\begin{figure}[h]
\center{\includegraphics[scale=0.3]{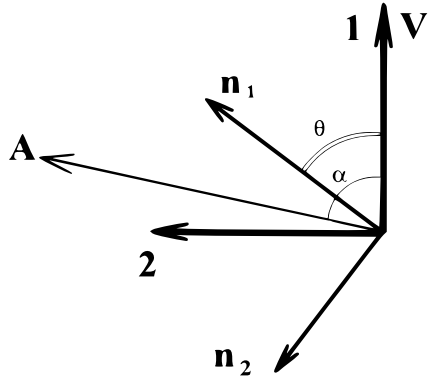}}
\caption{The coordinate system ($ \mathbf {n} ^ 1, \mathbf {n} ^ 2 $), in which the tensor $ s ^ {\alpha \beta} $ is diagonal. The vector $ \mathbf {n} ^ 3 $ is directed to the reader.}
\label{ris:1}
\end{figure}
 In this case, the components of the tensor $s ^ {\alpha \beta} $ are equal
\begin{equation}\label{7.01}
s^{11 }=\frac{VA\cos\alpha}{\sqrt{1-V^2}^{\,3}}
\end{equation}
\begin{equation}\label{7.02}
s^{12 }=s^{21 }=\frac{VA\sin\alpha}{2(1-V^2)}
\end{equation}
The remaining components of $s^{\alpha\beta}$ are zero. Solving the equation \eqref{27} we get the principal values of this tensor
\begin{equation}\label{7.03}
s^{1 }=\frac{VA}{2\sqrt{1-V^2}^{\,\,3}}\left(\sqrt{1-V^2\sin^2\alpha}+\cos\alpha\right)\,,
\end{equation}
\begin{equation}\label{7.04}
s^{2 }=-\frac{VA}{2\sqrt{1-V^2}^{\,\,3}}\left(\sqrt{1-V^2\sin^2\alpha}-\cos\alpha\right)\,.
\end{equation}
 The first principal direction, which corresponds to the value of $s^{1}$ is equal to
\begin{equation}\label{7.05}
\mathbf{n}^1=\frac{1}{\sqrt{2}}\left(\frac{\sqrt{\sqrt{1-V^2\sin^2\alpha}+\cos\alpha}}{\sqrt[4]{1-V^2\sin^2\alpha}}\,\,,\frac{\sqrt{\sqrt{1-V^2\sin^2\alpha}-\cos\alpha}}{\sqrt[4]{1-V^2\sin^2\alpha}}\,\,\right)\,.
\end{equation}
Accordingly, the second direction is
\begin{equation}\label{7.06}
\mathbf{n}^2=\frac{1}{\sqrt{2}}\left(-\frac{\sqrt{\sqrt{1-V^2\sin^2\alpha}-\cos\alpha}}{\sqrt[4]{1-V^2\sin^2\alpha}}\,\,,\frac{\sqrt{\sqrt{1-V^2\sin^2\alpha}+\cos\alpha}}{\sqrt[4]{1-V^2\sin^2\alpha}}\,\,\right)\,.
\end{equation}
We now find the orientation of the vector $\mathbf {n}^1$ relative to $\mathbf{V}$. The vector $\mathbf{n}^1$ can be represented as
\begin{equation}\label{7.07}
\mathbf{n}^1=\left(\cos\theta\,\,,\sin\theta\right)\,,
\end{equation}
where $ \theta $ is the angle of rotation $\mathbf {n}^1$. Comparing \eqref{7.07} with \eqref{7.06} we get that
\begin{equation}\label{7.08}
\tan^2\theta=\frac{\sqrt{1-V^2\sin^2\alpha}-\cos\alpha}{\sqrt{1-V^2\sin^2\alpha}+\cos\alpha}\,\,\,.
\end{equation}

\subsubsection*{4. The transformation of proper affine velocity into the laboratory reference frame.}

The inverse transformation to \eqref {52} has the form
\[\Omega ^{\alpha \beta } =\sqrt{1-V^{2} } \,\,\omega ^{\alpha \beta } -\frac{\sqrt{1-V^{2} } (1-\sqrt{1-V^{2} } )}{V^{2} }\, \omega ^{\gamma \beta } V^{\gamma } V^{\alpha } -\] 
\begin{equation} \label{a3.32} 
-\frac{(1-\sqrt{1-V^{2} } )^{2} }{V^{4} } \,(\omega ^{\gamma \mu } V^{\gamma } V^{\mu } )V^{\alpha } V^{\beta } +\frac{1-\sqrt{1-V^{2} } }{V^{2} }\, \omega ^{\alpha \gamma } V^{\beta } V^{\gamma } -\frac{V^{\beta } A^{\alpha } }{1-V^{2} }  \,.
\end{equation}
Here $ V^{\alpha}$ and $A^{\alpha}$ are the velocity and acceleration of the origin frame $k$, $\Omega^{\alpha\beta}$  and $ \omega^{\alpha \beta} $ is some matrix of affine velocity in the laboratory frame and in the nonrotating frame $k$, respectively. The calculation of this linear transformation is given in Appendix 3. We also indicate the transformation for $S^{\alpha \beta} $ and $ \Omega^ {\alpha} $.
\[S^{\alpha \beta } =\sqrt{1-V^{2} } \,\,s^{\alpha \beta } -\frac{(1-\sqrt{1-V^{2} } )^{2} }{V^{4} } (s^{\gamma \mu } V^{\gamma } V^{\mu } )V^{\alpha } V^{\beta }+\frac{(1-\sqrt{1-V^{2} } )^{2} }{2V^{2} } (s^{\alpha \gamma } V^{\beta } +s^{\beta \gamma } V^{\alpha } )V^{\gamma } -\] 
\begin{equation} \label{43} 
-\frac{1}{2} \left(e^{\gamma \lambda \beta } V^{\alpha } +e^{\gamma \lambda \alpha } V^{\beta } \right)\; \omega ^{\lambda } V^{\gamma }-\frac{V^{\beta } A^{\alpha } +V^{\alpha } A^{\beta } }{2(1-V^{2} )} \,,\end{equation}
 \begin{equation} \label{42}
 \Omega ^{\alpha } =\frac{2-V^2}{2}\,\,\,\omega ^{\alpha } -\frac{(1-\sqrt{1-V^{2} } )^{2} }{2V^{2} }\,\,(\boldsymbol{\omega}\mathbf{V})V^{\alpha} -\frac{1}{2}\, e^{\alpha \mu \nu } s^{\gamma \mu } V^{\nu } V^{\gamma }  -\frac{e^{\alpha \mu \nu } V^{\mu } A^{\nu } }{2(1-V^{2} )}\,. 
\end{equation}

Let us specially consider another important case of a rigid nonrotating local reference frame $k$ ($\omega ^ {\alpha \beta} = 0 $). From \eqref {a3.32} - \eqref {42} you can see that
\begin{equation} \label{a3.33} 
\Omega ^{\,\alpha \beta } =-\frac{V^{\beta } A^{\alpha } }{1-V^{2} }\,.
\end{equation} 
\begin{equation} \label{3.55} 
S^{\alpha \beta }=-\frac{V^{\beta } A^{\alpha } +V^{\alpha } A^{\beta } }{2(1-V^{2} )} \,,
\end{equation}
\begin{equation} \label{3.56}
 \mathbf{\Omega} =  -\frac{\mathbf{V\times A} }{2(1-V^{2} )}\,. 
\end{equation} 
Choose a coordinate system so that axis 1 passes along a bisectrix between the velocity vector $ \mathbf {V} $ and the acceleration vector $ \mathbf{A}$ (Figure 2).
\begin{figure}[ht]
\center{\includegraphics[scale=0.3]{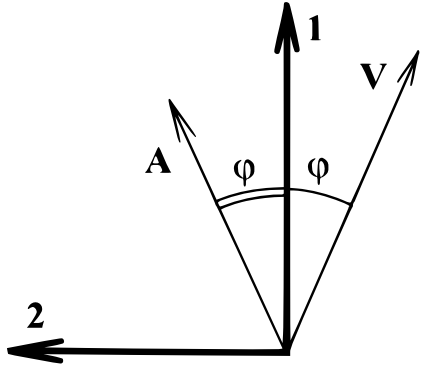}}
\caption{The coordinate system (\textbf {1,2}) in which the tensor $ S^{\alpha \beta}$ is diagonal.}
\label{ris:2}
\end{figure}
In this case, the tensor $S^{\alpha\beta}$ becomes diagonal. If the angle between the vectors $\mathbf {V}$ and $\mathbf{A}$ is $2\varphi$, then the principal values of this tensor are
\begin{equation}\label{3.58}
S^{1}=S^{11 }=-\frac{VA\,\cos^2\varphi}{2(1-V^2)}
\end{equation}
\begin{equation}\label{3.57}
S^{2}=S^{22 }=\frac{VA\,\sin^2\varphi}{2(1-V^2)}
\end{equation}
\begin{equation}\label{3.59}
S^{3}=S^{33 }=0
\end{equation}
For a reference frame that moves uniformly around a circumference with a frequency of $ \varpi $, the formula \eqref {3.56} goes to
\begin{equation} \label{a.3}
\mathbf{\Omega}= \mathbf{\Omega_{n}} =  -\frac{ V^2 }{2(1-V^{2} )}\,\boldsymbol{\varpi}=-\frac{\gamma^2-1}{2}\,\,\boldsymbol{\varpi}\,. 
\end{equation} 
It is interesting to compare this frequency value with the average value of the Thomas precession frequency, which in the case of uniform rotation is \cite {73}
\begin{equation} \label{a.4}
 \mathbf{\Omega}_{T} =  -\frac{ 1-\sqrt{1-{V^{}}^{2}} }{\sqrt{1-V^2}}\,\boldsymbol{\varpi}=-(\gamma-1)\,\boldsymbol{\varpi}\,. 
\end{equation} 
Then
\begin{equation} \label{a.5}
 \mathbf{\Omega_{n}} =\frac{\gamma+1}{2}\,\,\mathbf{\Omega}_{T}>\mathbf{\Omega}_{T} \,. 
\end{equation}

\subsubsection*{5. Angular acceleration disk.}
\label{fo}

Consider a rigidly rotating body, all points of which rotate around the origin of the laboratory inertial reference frame with velocity
\begin{equation} \label{57} 
\mathbf{V=\Omega \times R} \,.
\end{equation} 
Differentiating this equality, we obtain that the acceleration of $ \mathbf {A} $ is equal to
\begin{equation} \label{59} 
\mathbf{A=\Omega \times V+\dot{\Omega }\times R}\,. 
\end{equation} 
We select on this body a sufficiently small region near the point A with the coordinate $\mathbf {R}$. The rotational stiffness relative to the laboratory reference frame means the absence of the central stretching velocity of region A ($S^{\alpha \beta} = 0$). We will apply the previously obtained formulas to the nearest neighborhood of A.

We first consider the local deformation. The stretching rate along the main axis is determined by the tensor $s^{\alpha \beta}$ from the formula \eqref{56}. In this formula, the first three terms disappear due to the stiffness of the movement. Substituting here \eqref{59} as a result we get
\begin{equation} \label{62} 
s^{\alpha \beta } =\frac{e^{\alpha \delta \gamma } \dot{\Omega }^{\delta } R^{\gamma } V^{\beta } +V^{\alpha } e^{\beta \delta \gamma } \dot{\Omega }^{\delta } R^{\gamma } }{2(1-V^{2} )}  +\frac{1-\sqrt{1-V^{2} } }{V^{2} \sqrt{1-V^{2} } ^{\,\,3} } \left( \mathbf{ V  (\dot{\Omega }\times R})\right) \; V^{\alpha } V^{\beta } 
\end{equation} 
Thus, if the body rotates rigidly and uniformly with a constant angular velocity relative to the axial observer ($\mathbf{\dot{\Omega}} = 0$), then the strain rate tensor is $s^{\alpha\beta}$ in point A is zero. This means that the closest to the observer points of the body rotate without central stretching, that is, also as a rigid body. In the same case, if the body rotates with angular acceleration, the strain rate, generally speaking, is nonzero.

We are orient the axes of the coordinate system so that axis 2 is the direction of the angular velocity vector $ \mathbf {\Omega} $, vector $\mathbf{R}$ is the direction of axis 3, while the velocity vector $\mathbf {V = \Omega\times R} $ will be the direction of axis 1 (Figure 3).  
\begin{figure}[h]
\center{\includegraphics[scale=0.3]{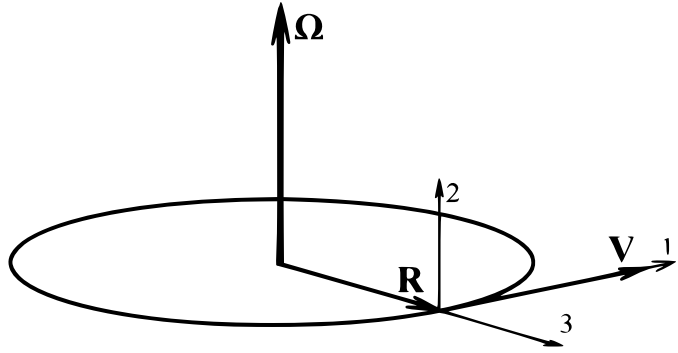}}
\caption{Local coordinate system on a rotating disk.}
\label{ris:3}
\end{figure}
Axes 1 and 3 in the laboratory frame change their direction in space. Let's move to the coordinate system that is connected with these directions. Then the tensor $s^{\alpha \beta}$ becomes diagonal and the only non-zero component is its principal value $s^{1}$ equal to \eqref{62}
\begin{equation} \label{63} 
s^{1} =s^{11} =\frac{V\dot{\Omega }R}{\sqrt{1-V^{2} } ^{\,\,3} } =\frac{R^{2} \Omega \dot{\Omega }}{\sqrt{1-R^{2} \Omega ^{2} } ^{\,\,3} } \,. 
\end{equation} 

We integrate this equality for the entire rotation from the initial state of rest to the angular velocity $\Omega$. We obtain that the logarithmic stretching factor along axis 1 is
\begin{equation} \label{66} 
\int\limits_{0}^{t}s^{1} dt = \int\limits_{0}^{T}\frac{R^{2} \Omega \dot{\Omega }}{\sqrt{1-R^{2} \Omega ^{2} } ^{\,\,3} }\cdot   \sqrt{1-R^{2} \Omega ^2}\,\,dT=-\frac{1}{2}\ln\left(1-R^2\Omega ^{2}\right)\,.
\end{equation} 
We are calculate the stretching of the body at point A according to the formula \eqref {25.3}. It is equal to 
\begin{equation} \label{4.63.1} 
r^{1}= \frac{r^{1}_0}{\sqrt{1-R^{2} \Omega ^{2}} }\,,                                                     
\end{equation}
\begin{equation} \label{67} 
r^{2}=r^{2}_0 \,,\,\,\,\,r^{3}=r^{3}_0\,.                                                     
\end{equation} 
This calculation agrees perfectly with the well-known fact that for a rotating disk during its accelerated rotation its material stretches in the direction of the axis 1 just $1/{\sqrt{1-R^{2}\Omega^{2}}}$ times .

Now consider the rotation.
 Since the rotation is rigid, the last term in \eqref {55} will disappear. As a result, we obtain in vector form, which is relative to the reference frame $k$
\begin{equation} \label{58} 
\boldsymbol{\omega} =\frac{2-V^2 }{2(1-V^{2}) }\,\mathbf{\Omega} -\frac{(1-\sqrt{1-V^{2} } )^{2} }{2V^2 (1-V^2)}\,\,  (\mathbf{\Omega V}) \mathbf{V} +\frac{\mathbf{V\times A}}{2(1-V^{2} )} \,. 
\end{equation} 
Substituting \eqref {59} into \eqref {58}, opened the brackets, leading similar terms, and considering that $\mathbf{\Omega V} = \mathbf {VR} = 0$ we get 
\begin{equation} \label{60} 
\boldsymbol{\omega} =\frac{\mathbf{\Omega}}{1-V^{2} } -\frac{\mathbf{R (\dot{\Omega }V)}}{2(1-V^{2} )}  \,.
\end{equation} 
 This formula gives the instantaneous angular velocity of the principal axes local region of rotating body. If the body spins along the axis of rotation, then $ \mathbf {\dot {\Omega} V} = 0 $, and the formula \eqref {60} is radically simplified
\begin{equation} \label{61} 
\boldsymbol{\omega} =\frac{\mathbf{\Omega}}{1-V^{2} }\,.  
\end{equation} 
This value is already known from various sources \cite {7}, \cite {41}.

\subsubsection*{Discussion}
\label{for}

Let's comment on the formulas now.

First of all, we note the characteristic properties of the generalized Lorentz transformation. Recall that this transformation describes the transition from the laboratory frame of reference to the frame $s$, which is experiencing Wigner's proper rotation. However, the parameter $\mathbf {v}$ of such a transformation  is not the velocity of the frame $s$, as one would expect. This parameter is the velocity of the frame $k$, which at the given time coincides with $s$, but does not rotate. This circumstance follows from the velocity addition formula given in Section 1 \eqref {a2.23}. This formula  relates the particle velocity in the $S$ and $k$ reference frames. Indeed, from the formula \eqref {a2.23} it follows that the points of the frame $k$ is rest in it ($\mathbf {u} = 0 $), only if they move relative to $ S $ with $ \mathbf {U = v} $.
 
 The equations \eqref {a2.15}, \eqref {a2.20} mean that the points of the coordinate system $ k $ moves at the velocity that is not equal to the velocity of the origin of the coordinates of the reference frame, and, therefore, moves heterogeneous from the point of view of the observer in $S$.  This circumstance is absolutely natural and expected.  Indeed, let the axis of the rectangular coordinate system of the accelerated reference frame be a rigid rod.  Then, in the process of its acceleration, with increasing speed, it will certainly be kinematically reduced relative to the laboratory frame.  This means that its points at a given moment of time move heterogeneous. The formula \eqref {a2.15} shows that during acceleration of the reference frame, which the origin is located at the rear end of a rigid (in its proper reference frame) rod, its front end moves slower than the rear end by $\mathbf{A(Vr)}/\sqrt {1-V ^ {2}}$.  In contrast, when braking, the front end moves faster than the rear end by the same amount.  If in classical mechanics the stiffness of the coordinate system assumes that the speed of movement of each of its points varies equally with time, regardless of its position, then this condition is not satisfied in relativistic theory.

This same fact can be looked at differently. Events occurring at the point with the coordinate $\mathbf {r} $ and in the origin of reference, simultaneous in $ s $ will be non-simultaneous in the laboratory frame.  The time interval between these events is equal to the desynchronization of the clock during the transition from the  reference frame $ s $ to the  laboratory frame $ S $. Thus, the inhomogeneity of motion of points of the coordinate system $ k $ is a consequence of the relativity of the notion of simultaneity in $ s $. In other words, there is actually a lag in the velocity increase of the rod  front point  compared to the back point by the time  $ \Delta T $  from \eqref {a2.70}.

The notion of heterogeneity in the movement of points of an accelerated rigid rod usually comes up against the objection that this velocity heterogeneity is incompatible with the rigidity of the rod. In other words, if we go to the instantly accompanying inertial frame of reference, moving at the velocity of the backsight point, then it would seem that the speed of the front point of the rod in this frame will be different from zero. It seems that if we assume the inhomogeneity of the velocities - the rigid coordinate system of the accelerated reference system will be absent. In fact, this objection is a prejudice based on the extension of the law of velocity subtraction (which is valid for inertial reference frames) for non-inertial frames.

The resulting affine velocity transformation was not simple. It is interesting to note that the equation \eqref {a3.32} consists of five terms, four of which exist even with uniform motion of $ k $, and the latter is a consequence of the influence of the acceleration of the  frame origin . On the contrary, the equation \eqref {52} consists of six terms, four of which exist even with a uniform motion of $ k $, and the last two are due to the effect of acceleration $ k $.  The complexity of this transformation may make it doubt its fairness. Nevertheless, this circumstance fully satisfies the requirements of the theory of relativity. In fact, the mathematical cumbersome transformation is a consequence of the diversity of the possible movement of points of a non-rigid reference frame in the laboratory frame.   From thes transformation it follows that, as a rule, the point of the reference frame moves not only rotating around the leading center with a some angular velocity, but also has a component of the velocity corresponding to stretching.

Let us consider important special cases of the obtained affine velocity transformation. Note that these cases were mentioned in the introduction to this article.
But then they managed to be considered in a uniform way, in one mathematical language.

The first of these particular cases was considered in Section 3 in example   of the transformation  \eqref {52}. This is a homogeneus motion of points, which Devan and Beran, and also Bell, considered for the first time. The importance of this case is determined by the fact that often the local frame of reference moves in such a way that the distance vector between any two points measured in $ S $ is constant during the movement. A simple example of such a motion is the motion of particles constituting the reference frame in a uniform force field. This means that in the \eqref {52} transformation, the  affine velocity in $S$ must be set equal to zero. It turned out that the distance between points in the $ k $ reference frame is increasing with the coefficients of relative elongation \eqref {7.03}, \eqref {7.04}. In this case, the principal axes of the system of points rotate in $ k $ with the rotation frequency \eqref {55.1}.

Another important case is the complex motion of the points of a rigid reference frame $ k $ in $ S $. This well-known problem is the subject of many papers \cite {2}. It is usually considered that points of $ k $ with respect to $S$ participate only in rotation. This phenomenon is called the Thomas precession. In fact, due to the Lorentz contraction, the movement of points $ k $ is complex. The only correct solution to the problem of this "precession"  was presented in the paper \cite[paragraph 3 and Appendix D]{16}. The author \cite {16}  (S. Stepanov)  obtained a differential equation for the length of a segment of a non-rotating reference system measured in a laboratory reference system. It has the form \eqref {255.1} (in our notation), where the affine velocity is \eqref {a3.33}. After that, Stepanov solved the corresponding differential equation and find the rod motion in the laboratory frame $ S $. The motion of the vector $ \mathbf {L} $ described by this equation turned out to be complicated and the angular velocity of rotation of the rod of the system $ k $ is variable. It is curious that the average Thomas frequency of the rod precession calculated by Stepanov  was the same as the generally accepted frequency of the Thomas precession. Thus, for the first time, he correctly set and solved the problem of the Thomas precession of a rigid rod.

The affine velocity  \eqref {a3.33} is the result of two movements. The first movement is the extension in $ S $ of the principal axes of the coordinate system $ k $ with the coefficient of relative elongation \eqref {3.58} - \eqref {3.59}. The second - is the rotation of the principal axes $ k $ with the frequency \eqref {3.56}. It was also calculated from the law of conservation of angular moment in \cite[formula (17)]{15}.

At first glance, the frequency of rotation of the principal axes of the $ k $ coordinate system should coincide with the frequency of the Thomas precession. However, from the formula \eqref {a.5} it follows that this is not the case. The frequency of the principal axes rotation with a uniform rotation in the frame $S$ in $(\gamma + 1) / 2$ times more than the frequency of the Thomas precession. This circumstance is connected with the fact that the principal axes  of the frame $ k $ is given by vectors external to $ k $: the velocity and the acceleration. Therefore, the principal axes is not the direction of some forever given rod of the frame $ k $. At successive moments of proper time, all new unit vectors $ k $ become the principal axes of the frame $ k $. The rods of the frame $ k $ of different directions rotate in the laboratory frame $ S $  non-uniformly.

It can be assumed that the neglect of the complex law of transformation of affine velocity apparently led to different formulas for the frequency of the Thomas precession.

The limit of applicability of the formulas obtained is basically reduced to the question of the degree of validity of the generalized Lorentz transformation. This transformation is almost certainly fair. The only limitation is that the proper acceleration of the reference frame origin should not experience jumps. In other words, the velocity parameter $\mathbf {v} $ in \eqref {1.1}, \eqref {1.2} must always be differentiable.

In the four-dimensional notation, the formulas obtained are difficult to present because of the non-covariance of the Lorentz contraction and time dilation. The meaning of the formulas obtained is apparently limited to a three-dimensional view. The reliability of the formulas obtained is confirmed by the standard method of the generalized Lorentz transformation adopted in this article and the coincidence of some results with the results of other authors.

\begin{flushleft}
{\bf{Summary.}}
\end{flushleft}

1. When considering non-inertial motion in the STR it is more expedient to use the method of generalized Lorentz transformation.

2. The  centro-affine motion of points in the local inertial reference frame $ k $ relative to the laboratory inertial reference frame $ S $ is also affine.

3.  Instead of the angular velocity of rotation of the body $ \mathbf {\Omega} $ in $ S $, only the affine velocity $ \Omega ^ {\alpha \beta} $ of the general non-antisymmetric form makes sense. Therefore, there is no Thomas precession (understood as pure rotation relative to the laboratory reference frame $ S $) for a body whose proper dimensions are preserved. By Thomas precession we mean the rotation of the principal axes of an accelerated moving body.

4. The resulting transformation of affine velocity is the result of the combined action of the basic effects of STR and is noncovariant, similar to other effects.

5. The proper local angular  velocity of a certain region of an arbitrarily rotating reference frame is equal to \eqref {60}.

6. The coincidence of the calculation of the local stretching of the spinning reference frame \eqref{4.63.1} with the already known value indirectly indicates the truth of the transformation of the affine velocity \eqref {a3.32}, \eqref{52}. This is also indirectly indicated by the coincidence of the calculated local proper angular velocity of a portion of a uniformly rotating rigid body with a known value \eqref{61}.

\begin{flushleft}
{\bf{Conclusion}}
\end{flushleft}

The novelty of this article is mainly in three aspects.

First, in this paper, an instantaneous velocity transformation \eqref {a2.23}, \eqref {a2.24} was found that connects the arbitrarily moving and non-rotating non-inertial frame $ k $ and the laboratory inertial frame $ S $. The physical meaning of the $\mathbf {v} \, (t) $  parameter of the LMN  transformation \eqref {1.1}, \eqref {1.2} was also shown. Remarkably, it turned out to be simple.  The parameter $ \mathbf {v} \, (t) $ is the velocity of the points of the non-rotating reference frame $ k $, which accompanies  the frame $ s $. The motion of $ k $ is non-uniform, according to the formula \eqref {a2.15}. The points of the frame $ s $ move relative to $ S $ with the velocity \eqref {a2.14}.

 In addition, a transformation of affine velocity from the frame $ k $ into the frame $ S $ and vice versa is found. The obtained formulas turned out to be rather cumbersome, but in practice it is not the affine velocity of a small region that is interesting, but its angular velocity of rotation of the principal axes and the velocity of its stretching along the principal axes. For these quantities, the transformation laws \eqref {55}, \eqref {56} were also found. Important special cases of this transformation are also considered.

Finally, the equations obtained were applied to an accelerated rotating reference frame and its local angular velocity $ \boldsymbol {\omega} $ was calculated.  It turned out that if the rotation of the body relative to the observer in the laboratory frame was rigid, then when the observer is at the periphery of the body, body moves, generally speaking, non-rigid \eqref {62}. If the distance between two points relative to the observer in the center of an unevenly rotating body is maintained, then when the observer shifts, this distance will change even in the near vicinity of the observer. It turned out that the value of this angular  velocity  $ \boldsymbol {\omega} $ also depends on the angular acceleration of the reference frame. The angular velocity of rotation of points in the nearest vicinity around the observer for the case of nonuniform rotation will be equal to \eqref {60}. For a rigid body, the proper extension of the periphery \eqref {4.63.1} turned out to be the same as the known value.
	
The practical value of the known dependence of affine velocity as a function of time for kinematics is large. It is enough to say that her knowledge in principle allows us to solve a differential equation of the form \eqref {2.25}, just as Stepanov decided, and thus, to find the motion of a deformable rod in a laboratory frame of reference. Therefore, the known dependence of affine velocity on time for kinematics plays the role of force for the 2 Newton's law. Accordingly, the found transformation of affine velocity in kinematics takes the place of the formula for the transformation of forces in dynamics.

There is another far-reaching analogy: between the kinematic characteristics and the field strength. Indeed, the group acceleration of particles is in some way analogous to the strength of an electric field, and the angular velocity $ \mathbf {\Omega} $ analogous to  the magnetic field strength. On the other hand, we now know that the 3 components of the angular velocity vector $ \mathbf {\Omega} $ and the 6 components of the symmetric tensor $ S ^ {\alpha \beta} $ together form one affine velocity tensor $ \Omega ^ {\alpha \beta} $.  The question arises: is there also a new, third field (except electric and magnetic) of a tensor type of a quasi-magnetic nature, which in relation to its effect on a group of test charges leads only to a  size change   (the stretching or the contraction)  of the group (similarly to $ S ^ {\alpha \beta} $)? This gives reason to believe known transformation law  of electromagnetic fields when changing the reference frame  is not quite correct. We believe that in the derivation of the future correct relativistic law of transformation of field strengths, the transformation of affine velocity will play a decisive role.

\newpage

\subsubsection*{Appendix 1. The velocity in proper and laboratory reference frames.} 
\label{first}

Differentiating the relations \eqref {1.1}, \eqref {1.2} we get
\begin{equation} \label{2.5} 
dT=\left\{\frac{1}{\sqrt{1-v^{2} } } +\frac{\mathbf{\dot{v}r}}{\sqrt{1-v^{2} } } +\frac{(\mathbf{\dot{v}v)(vr)}}{\sqrt{1-v^{2} } ^{3} } \right\}dt+\frac{\mathbf{v}d\mathbf{r}}{\sqrt{1-v^{2} } }  
\end{equation} 
\[d\mathbf{R}=\left\{\frac{2\sqrt{1-v^{2} } ^{\,\,3} +3v^{2} -2}{v^{4} \sqrt{1-v^{2} } ^{\,\,3} } \mathbf{(vr)}(\mathbf{\dot{v}v)v}+\frac{1-\sqrt{1-v^{2} } }{v^{2} \sqrt{1-v^{2} } } \left[\mathbf{(vr)\dot{v}+(\dot{v}r)v}\right]+\frac{\mathbf{v}}{\sqrt{1-v^{2} } } \right\}dt\] 
\begin{equation} \label{2.6} 
+\,d\mathbf{r}+\frac{1-\sqrt{1-v^{2} } }{v^{2} \sqrt{1-v^{2} } } (\mathbf{v}d\mathbf{r)v} 
\end{equation} 	

From \eqref {3} it follows that
\begin{equation} \label{a2.7} 
\dot{\mathbf{v}}=\sqrt{1-v^{2} }\, \mathbf{W}-\frac{\sqrt{1-v^{2} } (1-\sqrt{1-v^{2} } )}{v^{2} } \mathbf{(vW)v} 
\end{equation} 

Substituting this relation into \eqref {4} we get
\begin{equation} \label{a2.8} 
\mathbf{\Omega} _{W} =\frac{1-\sqrt{1-v^{2} } }{v^{2} } \,\mathbf{v\times W} 
\end{equation} 

Given \eqref {a2.7}, \eqref {a2.8}, the formulas \eqref {2.5}, \eqref {2.6} can be rewritten in a more compact form
\begin{equation} \label{a2.9} 
dT=\frac{1+(\mathbf{W+v\times \Omega} _{W} )\,\mathbf{r}}{\sqrt{1-v^{2} } }\, dt+\frac{\mathbf{v}d\mathbf{r}}{\sqrt{1-v^{2} } }  
\end{equation} 
\begin{equation} \label{a2.10} 
d\mathbf{R}=\left\{\frac{1+\mathbf{Wr}}{\sqrt{1-v^{2} } }\, \mathbf{v}\right. +\mathbf{\Omega} _{W} \times \mathbf{r}+\left. \frac{1-\sqrt{1-v^{2} } }{v^{2} \sqrt{1-v^{2} } } \left[\mathbf{v}\, (\mathbf{\Omega} _{W} \times \mathbf{r})\right]\;\mathbf{v}\right\}dt+d\mathbf{r}+\frac{1-\sqrt{1-v^{2} } }{v^{2} \sqrt{1-v^{2} } }\, (\mathbf{v}d\mathbf{r)v} 
\end{equation} 

We divide \eqref {2.6} by \eqref {2.5} and enter the velocityes $ \mathbf {U} $ and $ \mathbf {u} $ 
\[\mathbf{U}=\frac{d\mathbf{R}}{dT}\, ,\,\,\,        \mathbf{u}_{s}=\frac{d\mathbf{r}}{dt} \] 
in the reference frames $ S $ and $ s $. They are related by the equation
 \begin{equation} \label{a2.11} 
\mathbf{U}=\frac{(1+\mathbf{Wr)v}+\sqrt{1-v^{2} } (\mathbf{u}_{s}+\mathbf{\Omega} _{W} \times\mathbf{r})+\frac{1-\sqrt{1-v^{2} } }{v^{2} } \left[\mathbf{v}\; (\mathbf{u}_{s}+\mathbf{\Omega} _{W} \times \mathbf{r})\right]\; \mathbf{v}}{1+\mathbf{Wr+v\; (u}_{s}+\mathbf{\Omega} _{W} \times \mathbf{r})} .                  
\end{equation} 
Reversing this equality, we get
\begin{equation} \label{a2.12} 
\mathbf{u}_{s}=\frac{(1+\mathbf{Wr})\; \left[\sqrt{1-v^{2} }\,\mathbf{U-v}+\frac{1-\sqrt{1-v^{2} } }{v^{2} } \mathbf{v(vU})\right]}{1-\mathbf{vU}} -\mathbf{\Omega} _{W} \times \mathbf{r} 
\end{equation} 

  Note that in the formulas \eqref {a2.11} and \eqref {a2.12} the value $ \mathbf {u} _ {s} + \mathbf {\Omega} _ {W} \times \mathbf {r} $ is the velocity $\mathbf {u} $ of a point in the non-rotating reference frame $ k $ that accompanies the frame   $ s $ . Then these formulas are simplified and have the form \eqref {a2.23}, \eqref {a2.24}.

\subsubsection*{Appendix 2. The velocity of the reference frame $ s $.}

From \eqref {a2.11} it can be seen that in order for the points of the 3-space of the frame $ s $ to rest relative to its coordinate system ($ \mathbf {u} _ {s} = 0 $) it is necessary that they move relative to $ S $ not with velocity $ \mathbf {U = v} $, but with velocity $ \mathbf {U = U} _ {0} $
\begin{equation} \label{a2.13} 
\mathbf{U}_{0} =\frac{(1+\mathbf{Wr)v}+\sqrt{1-v^{2} }\,\, \mathbf{\Omega} _{W} \times \mathbf{r}+\frac{1-\sqrt{1-v^{2} } }{v^{2} } \left[\mathbf{v\; (\Omega} _{W} \times \mathbf{r})\right]\; \mathbf{v}}{1+(\mathbf{W+v\times \Omega} _{W} )\; \mathbf{r}} \,. 
\end{equation} 
This shows that the value of $ \mathbf {v} $ is not the velocity of the points of the reference frame $ s $. Expanding this expression in powers of $ \mathbf {r} $ in the first approximation, we get
\begin{equation} \label{a2.14} 
\mathbf{U}_{0} =\mathbf{v}+\sqrt{1-v^{2} }\,\,\mathbf{ \Omega} _{W} \times \mathbf{r}-\frac{\sqrt{1-v^{2} } \,(1-\sqrt{1-v^{2} } )}{v^{2} } \left[\mathbf{v(\Omega} _{W} \times \mathbf{r})\right]\mathbf{v}\,. 
\end{equation} 

We will take into account that
\begin{equation} \label{a2.16} 
\mathbf{\dot{v}}=\frac{1}{\sqrt{1-V^{2} } } \frac{d\,\mathbf{V}}{dT} \,.                                                  
\end{equation} 
Then the proper acceleration and angular velocity will be equal
\begin{equation} \label{a2.17} 
\mathbf{W}=\frac{\mathbf{\dot{V}}}{1-V^{2} } +\frac{1-\sqrt{1-V^{2} } }{V^{2} \sqrt{1-V^{2} } ^{\,\,3} }\mathbf{(V\dot{V})V}\,, 
\end{equation} 
\begin{equation} \label{a2.18} 
\mathbf{\Omega}_{W} =\frac{1-\sqrt{1-V^{2} } }{V^{2} (1-V^{2} )}\mathbf{V\times \dot{V}}, \,\,\,\,              \mathbf{\dot{V}}=\frac{d\mathbf{V}}{dT}  \,.
\end{equation} 

From \eqref {a2.17} we can see that
\begin{equation} \label{a2.19} 
\mathbf{\dot{V}}=(1-V^{2} )\mathbf{W}-\frac{(1-V^{2} )(1-\sqrt{1-V^{2} } )}{V^{2} } \mathbf{(VW)V} \,.
\end{equation} 
Therefore, substituting \eqref {a2.19} in \eqref {a2.15} we get\eqref{a2.20}, where $ \mathbf {V} $ is the speed of origin of $ s $ relative to $ S $. Substituting in \eqref {a2.14} the value \eqref {a2.20} we get
\[\mathbf{U}_{0} =\mathbf{V}-\sqrt{1-V^{2} }\mathbf{(Vr)W}+\frac{\sqrt{1-V^{2} } (1-\sqrt{1-V^{2} } )}{V^{2} } \mathbf{(Vr)(VW)V}+\] 
\begin{equation} \label{a2.21} 
+\sqrt{1-V^{2} } \mathbf{\Omega} _{W} \times\mathbf{r}-\frac{\sqrt{1-V^{2} } (1-\sqrt{1-V^{2} } )}{V^{2} } \left[\mathbf{V(\Omega} _{W} \times \mathbf{r})\right]\mathbf{V} \,.
\end{equation} 
Substituting here \eqref {a2.17}, \eqref {a2.18} we get
\[\mathbf{U}_{0} =\mathbf{V}-\frac{1-\sqrt{1-V^{2} } }{V^{2} } \mathbf{(Vr)\dot{V}}-\frac{(1-\sqrt{1-V^{2} } )^{2} }{V^{4} \sqrt{1-V^{2} } } \mathbf{(Vr)(\dot{V}V)V}-\]
\begin{equation} \label{a2.22} 
-\frac{1-\sqrt{1-V^{2} } }{V^{2} } \mathbf{(\dot{V}r)V} \,.
\end{equation}

\subsubsection*{Appendix 3. The transformation of affine velocity into a comoving non-rotating reference frame.}
\label{thirth}

What is the dependence of $ \omega ^ {\alpha \beta} $ on $ \Omega ^ {\alpha \beta} $ according to the theory of relativity? We derive it from the velocity subtraction formula \eqref {a2.24}. 
We substitute in \eqref {a2.24}  the expression \eqref {255} for small $ \mathbf {r} $ and the relation \eqref {a2.15} between the transformation function $ \mathbf {v} (t) $ and the velocity  $ \mathbf {V} $ the coordinates origin $ s $.
The answer must be sought with the accuracy of the terms proportional to the first degree of the distance. Note that the second  multiplier of the numerator of the right side \eqref {a2.24} is already proportional to the component $ r ^ {\alpha} $, therefore, up to the first degrees of $ r ^ {\alpha} $, the first multiplier can be set equal to 1, and the denominator is $ 1-V ^ {2} $. Further, using \eqref {a2.15} it is easy to calculate that
\begin{equation} \label{49} 
\sqrt{1-v^{2} } =\sqrt{1-V^{2} } +\frac{(\mathbf{\dot{V}V)(Vr)}}{1-V^{2} } \,, 
\end{equation} 
\[\frac{1-\sqrt{1-v^{2} } }{v^{2} }\, v^{\alpha } v^{\beta } =\frac{1-\sqrt{1-V^{2} } }{V^{2} }\, V^{\alpha } V^{\beta } -\frac{(1-\sqrt{1-V^{2} } )^{2} }{V^{4} (1-V^{2} )} (\mathbf{V\dot{V})(Vr)}V^{\alpha } V^{\beta } -\] 
\begin{equation} \label{50} 
-\frac{1-\sqrt{1-V^{2} } }{V^{2} \sqrt{1-V^{2} } } \mathbf{(Vr)}(\dot{V}^{\alpha } V^{\beta } +\dot{V}^{\beta } V^{\alpha } )\,. 
\end{equation} 

Given these considerations, it can be found that

\[u^{\alpha } =\frac{1-\sqrt{1-V^{2} } }{V^{2} \sqrt{1-V^{2} } ^{\,3} } (\mathbf{\dot{V}V)(Vr)}\, V^{\alpha } +\frac{\mathbf{(Vr)}}{1-V^{2} } \dot{V}^{\alpha } +\frac{1}{\sqrt{1-V^{2} } } \,\Omega ^{\alpha \beta } L^{\beta }+\] 
 \begin{equation} \label{51} 
+\frac{1-\sqrt{1-V^{2} } }{V^{2} (1-V^{2} )} (\Omega ^{\beta \gamma } V^{\beta } L^{\gamma } )V^{\alpha }\,,  
\end{equation} 
where $ L ^ {\ alpha} $ is defined by the Lorentz contraction formula.
In the first order, the distance $ \mathbf {L} $ from the point to the origin of the system $ k $, which see relative to $ S $ equal to
\begin{equation} \label{a3.27} 
\mathbf{L=r}-\frac{1-\sqrt{1-V^{2} } }{V^{2} }\, \mathbf{(rV)V}\,.                                              
\end{equation} 
Substituting formula \eqref {a3.27} in \eqref {51} and comparing the resulting expression with \eqref {a2.25}, we can obtain the equality \eqref {52}.

From \eqref {52}, one can find the angular velocity $ \omega ^ {\alpha} $ of the principal axes of the affine tensor \eqref {2.40} and the central tension velocity tensor $s ^ {\alpha \beta} $ \eqref {2.41}. Calculations give
\[\omega ^{\alpha } =\frac{e^{\alpha \mu \nu } (\Omega ^{\nu \mu } -\Omega ^{\mu \nu } )}{4\sqrt{1-V^{2} } } -\frac{1-\sqrt{1-V^{2} } }{4V^{2} \sqrt{1-V^{2} } } \,\,e^{\alpha \mu \nu } V^{\gamma } (\Omega ^{\nu \gamma } V^{\mu } -\Omega ^{\mu \gamma } V^{\nu } )+\] 
\begin{equation} \label{53} 
+\frac{1-\sqrt{1-V^{2} } }{4V^{2} (1-V^{2} )} e^{\alpha \mu \nu } V^{\gamma } (\Omega ^{\gamma \mu } V^{\nu } -\Omega ^{\gamma \nu } V^{\mu } )+\frac{e^{\alpha \mu \nu } }{2(1-V^{2} )} V^{\mu } \dot{V}^{\nu } \,, 
\end{equation} 
\[s^{\alpha \beta } =\frac{\Omega ^{\alpha \beta } +\Omega ^{\beta \alpha } }{2\sqrt{1-V^{2} } } - \frac{1-\sqrt{1-V^{2} } }{2V^{2} \sqrt{1-V^{2} } } \,V^{\gamma } (\Omega ^{\alpha \gamma } V^{\beta } 
+\Omega ^{\beta \gamma } V^{\alpha } )+\frac{1-\sqrt{1-V^{2} } }{2V^{2} (1-V^{2} )} \,V^{\gamma } (\Omega ^{\gamma \beta } V^{\alpha } +\Omega ^{\gamma \alpha } V^{\beta } )-\]
\begin{equation} \label{54}
-\frac{(1-\sqrt{1-V^{2} } )^{2} }{V^{4} (1-V^{2} )} (\Omega ^{\gamma \mu } V^{\gamma } V^{\mu } )V^{\alpha } V^{\beta}+\frac{1-\sqrt{1-V^{2} } }{V^{2} \sqrt{1-V^{2} } ^{\,\,3} } (\mathbf{V\dot{V}})V^{\alpha } V^{\beta } +\frac{\dot{V}^{\alpha } V^{\beta } +V^{\alpha } \dot{V}^{\beta } }{2(1-V^{2} )}\, .       
\end{equation} 

 Similarly \eqref {2.39} - \eqref {2.41}, we introduce the  symmetric tensor $ S ^ {\alpha \beta} $ and the vector $ \Omega ^ {\alpha}$, which is dual to the antisymmetric part $ \Omega ^ {\alpha \beta} $.  
\begin{equation} \label{b3.34} 
\Omega ^{\alpha } =\frac{1}{4} \,e^{\alpha \mu \nu } \left(\Omega ^{\nu \mu } -\Omega ^{\mu \nu } \right)\,,       
\end{equation} 
\begin{equation} \label{b3.35} 
S^{\alpha \beta } =\frac{1}{2} \left(\Omega ^{\alpha \beta } +\Omega ^{\beta \alpha } \right)\,.                                               
\end{equation}
 Therefore
\begin{equation} \label{b3.36} 
\Omega ^{\alpha \beta } =S^{\alpha \beta } +e^{\alpha \gamma \beta } \Omega ^{\gamma } \,. \end{equation}
Then the formula \eqref {52} can be somewhat simplified. Using the equations \eqref {b3.34} - \eqref {b3.36} we find the formulas \eqref {56} and \eqref {55}.

\subsubsection*{Appendix 4. The transformation of  proper affine velocity into a laboratory reference frame}
\label{sixth}

The dependence of $ \omega ^ {\alpha \beta} $ on $ \Omega ^ {\alpha \beta} $ could be found directly from \eqref {52}. We will proceed differently and derive it from \eqref {a2.24}. We derive it from the velocity addition formula \eqref {a2.23}. It is obvious that in the second and third terms of the numerator and in the denominator with the required accuracy $ \mathbf {v} $ can be replaced by $ \mathbf {V} $. Then, substituting the formula \eqref {2.25} in \eqref {a2.23} and in the first term of the numerator, \eqref {a2.15}, we obtain by expanding the fraction up to the first powers of $\mathbf{r}$
\begin{equation} \label{a3.28} 
U^{\alpha } =V^{\alpha } -\frac{\mathbf{(Vr)}\, \dot{V}^{\alpha } }{\sqrt{1-V^{2} } } +\sqrt{1-V^{2} }\,\, \omega ^{\alpha \beta } r^{\beta } -\frac{\sqrt{1-V^{2} } (1-\sqrt{1-V^{2} } )}{V^{2} }\, \omega ^{\gamma \beta } V^{\gamma } V^{\alpha } r^{\beta }  
\end{equation}

We substitute in \eqref {a3.28}  the expression is the opposite of the Lorentz contraction formula instead of $ \mathbf {r} $
\begin{equation} \label{a3.29} 
\mathbf{r=L}+\frac{1-\sqrt{1-V^{2} } }{V^{2} \sqrt{1-V^{2} } } \mathbf{(LV)V}.                                                         
\end{equation} 
From here    
\[U^{\alpha } =V^{\alpha } +\sqrt{1-V^{2} }\,\, \omega ^{\alpha \beta } L^{\beta }-\frac{\sqrt{1-V^{2} } (1-\sqrt{1-V^{2} } )}{V^{2} } (\omega ^{\gamma \beta } V^{\gamma } L^{\beta } )V^{\alpha } +\] 
\begin{equation} \label{a3.30} 
+\frac{1-\sqrt{1-V^{2} } }{V^{2} } \mathbf{(VL)}\, \omega ^{\alpha \gamma } V^{\gamma } -\frac{(1-\sqrt{1-V^{2} } )^{2} }{V^{4} } \,\mathbf{(VL)}(\, \omega ^{\gamma \mu } V^{\gamma } V^{\mu } )V^{\alpha } -\frac{\mathbf{(VL)}\dot{V}^{\alpha } }{1-V^{2} }  
\end{equation} 
The expression \ eqref {a3.30} cannot be represented as
\begin{equation} \label{a3.31} 
U^{\alpha } =V^{\alpha } +\dot{L}^{\alpha }=V^{\alpha } +e^{\alpha \beta \gamma } \Omega ^{\beta } L^{\gamma } ,   
\end{equation} 
with some $ \Omega ^ {\alpha} $, it means that the affine motion of points in the frame $ k $ relative to the laboratory frame $ S $ is not rigid. Comparing \eqref {a3.30} with \eqref {255} we finally get the equality \eqref {a3.32}.

 The formula \eqref {a3.32} can be somewhat simplified by introducing, similarly to \eqref {2.39} - \eqref {2.41}, the symmetric tensor $S^ {\alpha \beta} $ and the vector $\Omega^{\alpha}$, which is dual of antisymmetric part $ \Omega^{\alpha \beta} $. 
\begin{equation} \label{a3.34} \Omega ^{\alpha } =\frac{1}{4} \,e^{\alpha \mu \nu } \left(\Omega ^{\nu \mu } -\Omega ^{\mu \nu } \right)\,,       \end{equation} \begin{equation} \label{a3.35} S^{\alpha \beta } =\frac{1}{2} \left(\Omega ^{\alpha \beta } +\Omega ^{\beta \alpha } \right)\,.                                               \end{equation}
Therefore
\begin{equation} \label{a3.36} \Omega ^{\alpha \beta } =S^{\alpha \beta } +e^{\alpha \gamma \beta } \Omega ^{\gamma } \,. \end{equation} 
Substitution \eqref {a3.34} - \eqref {a3.36} into \eqref {a3.32} gives
\[\Omega ^{\alpha } =\frac{\sqrt{1-V^{2} }\,\, e^{\alpha \mu \nu } (\omega ^{\nu \mu } -\omega ^{\mu \nu } )}{4} -\frac{\sqrt{1-V^{2} } (1-\sqrt{1-V^{2} } )}{4V^{2} }\, e^{\alpha \mu \nu } (\omega ^{\gamma \mu } V^{\nu } -\omega ^{\gamma \nu } V^{\mu } )V^{\gamma } +\] 
\begin{equation} \label{a3.37} +\frac{1-\sqrt{1-V^{2} } }{4V^{2} } \,e^{\alpha \mu \nu } (\omega ^{\nu \gamma } V^{\mu } -\omega ^{\mu \gamma } V^{\nu } )V^{\gamma }  -\frac{e^{\alpha \mu \nu } }{4(1-V^{2} )} (V^{\mu } \dot{V}^{\nu } -V^{\nu } \dot{V}^{\mu } ) \,,\end{equation} 
\[S^{\alpha \beta } =\frac{\sqrt{1-V^{2} } }{2} (\omega ^{\alpha \beta } +\omega ^{\beta \alpha } )-\frac{\sqrt{1-V^{2} } (1-\sqrt{1-V^{2} } )}{2V^{2} } (\omega ^{\gamma \beta } V^{\alpha } +\omega ^{\gamma \alpha } V^{\beta } )V^{\gamma } -\] 
\begin{equation} \label{a3.38} -\frac{(1-\sqrt{1-V^{2} } )^{2} }{V^{4} } (\omega ^{\gamma \mu } V^{\gamma } V^{\mu } )V^{\alpha } V^{\beta } +\frac{1-\sqrt{1-V^{2} } }{2V^{2} } (\omega ^{\alpha \gamma } V^{\beta } +\omega ^{\beta \gamma } V^{\alpha } )V^{\gamma } -\frac{V^{\beta } \dot{V}^{\alpha } +V^{\alpha } \dot{V}^{\beta } }{2\sqrt{1-V^{2} } } \,. \end{equation} 
Given \eqref {2.39} - \eqref {2.41}, we get \eqref {42} and \eqref {43}.

\vspace{20pt}

\small

\makeatletter
\@addtoreset{equation}{section}
\@addtoreset{footnote}{section}
\renewcommand{\section}{\@startsection{section}{1}{0pt}{1.3ex
plus 1ex minus 1ex}{1.3ex plus .1ex}{}}

{ 

\renewcommand{\refname}{{\rm}

\end{document}